\documentclass{aa}
\usepackage{graphicx}
\usepackage{txfonts}
\usepackage{natbib}
\begin{document}
\title{Electron temperature fluctuations in NGC\,346}

\author{V.~A. Oliveira\inst{1} \and M.~V.~F. Copetti\inst{1} \and A.~C. Krabbe\inst{2}}   

\offprints{Vinicius de A. Oliveira, \\
\email{vinic99@mail.ufsm.br} }
\titlerunning{Electron temperature fluctuations in NGC\,346}
\institute{
    Laborat\'orio de An\'alise Num\'erica e Astrof\'{\i}sica, Departamento de Matem\'atica, \\
    Universidade Federal de Santa Maria, 97119-900 Santa Maria, RS, Brazil. \\
  \and Southern Astrophysical Research Telescope, c/o AURA Inc., Casilla 603, La Serena, Chile. \\} 
\date{Received July 8, 2008 / Accepted September 28, 2008}


\abstract
{
The existence and origin of large spatial temperature fluctuations in \ion{H}{ii} regions and planetary nebulae are assumed to explain the differences between the heavy element abundances inferred from collisionally excited and recombination lines, although this interpretation remains significantly controversial.
}
{
We investigate the spatial variation in electron temperature inside \object{NGC\,346}, the brightest \ion{H}{ii} region in the Small Magellanic Cloud.
}
{
Long slit spectrophotometric data of high signal-to-noise were employed to derive the electron temperature from measurements derived from localized observations of the [\ion{O}{iii}]($\lambda4959 + \lambda5007)/\lambda4363$ ratio in three directions across the nebula. 
}
{
The electron temperature was estimated in 179 areas of 5\arcsec$\times$1\farcs5 of size distributed along three different declinations. A largely homogeneous temperature distribution was found with a mean temperature of 12\,269 K and a dispersion of 6.1\%. After correcting for pure measurements errors, a temperature fluctuation on the plane of the sky of $t^2_{\rm s} = 0.0021$ (corresponding to a dispersion of 4.5\%) was obtained, which indicates a 3D temperature fluctuation parameter of $t^2 \approx 0.008$. A large scale gradient in temperature of the order of $-5.7\pm1.3$ K arcsec$^{-1}$ was found. 
}
{
The magnitude of the temperature fluctuations observed agrees with the large scale variations in temperature predicted by standard photoionization models, but is too small to explain the abundance discrepancy problem. However, the possible existence of small spatial scale temperature variations is not excluded.
}
\maketitle

 %

\section{Introduction}
%

The most important unsolved issue in the study of photoionized nebulae is why the chemical abundances of heavy elements derived from recombination lines are systematically higher than those derived from collisionally excited lines, the so-called ``abundance discrepancy" (AD) problem. For \ion{H}{ii} regions, the ratio between the abundances estimated from permitted and forbidden lines is about ${\rm ADF} \approx 2$ \citep{Garcia-Rojas & Esteban 2007}. Much larger abundance discrepancies ($2 \lesssim {\rm ADF} \lesssim 20$) have been found in planetary nebulae \citep{Liu et al. 2000, Liu et al. 2001}. Spatial temperature fluctuations have been proposed to explain the AD problem. These temperature fluctuations were initially proposed by \citet{Peimbert 1967} to account for the considerable discrepancy found between the temperature estimates for \ion{H}{ii} regions obtained by different methods. However, the high levels of temperature fluctuation required to solve the AD problem are not predicted by standard photoionisation models \citep{Kingdon & Ferland 1995}. Chemical or density inhomogeneities of different characteristics and origins have been proposed to explain the high temperature fluctuations needed \citep{Giammanco & Beckman 2005, Kingdon & Ferland 1998, Liu et al. 2000, Stasinska et al. 2007, Tsamis et al. 2004, Viegas & Clegg 1994}, but the results of analysis based on observations are still inconclusive. 

Direct determinations of electron temperature fluctuations in the plane of the sky have been obtained for about a dozen of planetary nebulae \citep{Krabbe & Copetti 2005, Liu 1998, Rubin et al. 2002, Tsamis et al. 2008} and two \ion{H}{ii} regions, the \object{Orion Nebula} \citep{Mesa-Delgado et al. 2008, O'Dell et al. 2003, Rubin et al. 2003} and the \object{30 Doradus Nebula} \citep{Krabbe & Copetti 2002}. Localized measurements of electron temperature were obtained across the object surface area and small temperature fluctuations were inferred.

In this paper, we describe a study of spatial variations in electron temperature within \object{NGC\,346}, the brightest \ion{H}{ii} region in the Small Magellanic Cloud (SMC). This object is also designated as \object{LHA 115-N\,66} in the catalogue of emission nebulae in the Magellanic Clouds by \citet{Henize 1956} and \object{DEM\,S\,103} in that of \citet{Davies et al. 1976}, or more concisely as \object{N\,66} and \object{DEM\,103}, respectively. \object{NGC\,346} is a 14\arcmin$\times$11\arcmin\ nebula with complex structure, composed of arcs, filaments, and compact blobs of about  30\arcsec\ or less (e.g. the component \object{N\,66A}). The temperatures were derived from localized measurements of the [\ion{O}{iii}]($\lambda4959 + \lambda5007)/\lambda4363$ emission line ratio obtained from long-slit spectrophotometric observations of high signal-to-noise ratio.

\section{Observations and data reduction}
%

\begin{table*}[t]
\caption{[\ion{O}{iii}] ratio and electron temperature statistics}
\label{tab:estat}
\begin{tabular}{lllllcllll}
\hline
\hline
\noalign{\smallskip}
 & \multicolumn{4}{c}{[\ion{O}{iii}] ratio} & & \multicolumn{4}{c}{$T_{\rm e}$ (K)}\\ 
\noalign{\smallskip}
\cline{2-5} \cline{7-10}
\noalign{\smallskip}
 & \multicolumn{4}{c}{Slit position $\Delta\delta$} & & \multicolumn{4}{c}{Slit position $\Delta\delta$} \\
 & 30\arcsec N & 0\arcsec & 30\arcsec S & all & & 30\arcsec N & 0\arcsec & 30\arcsec S & all \\
\hline
\noalign{\smallskip}
number of data $N$ 	    & 63  & 58  & 58  & 179 & & 63    & 58    & 58    & 179   \\
minimum 	     	    & 58  & 67  & 96  & 58  & & 10082 & 9598  & 9078  & 9078  \\
first quartile $Q1$  	    & 102 & 94  & 120 & 99  & & 12140 & 12239 & 11384 & 11786 \\
median 	             	    & 112 & 103 & 130 & 111 & & 12581 & 12695 & 11657 & 12331 \\
third quartile $Q3$  	    & 118 & 114 & 138 & 126 & & 12766 & 13174 & 12235 & 12879 \\
maximum 		    & 203 & 239 & 168 & 291 & & 16291 & 15142 & 14481 & 16291 \\
weighted mean  	 	    & 112 & 105 & 131 & 115 & & 12418 & 12678 & 11713 & 12269 \\
weighted standard deviation & 9   & 10  & 11  & 13  & & 635   & 675   & 591   & 751   \\
\noalign{\smallskip}
\hline
\noalign{\smallskip}
\end{tabular}           
\end{table*}

\begin{figure}
\resizebox{\hsize}{!}{\includegraphics*{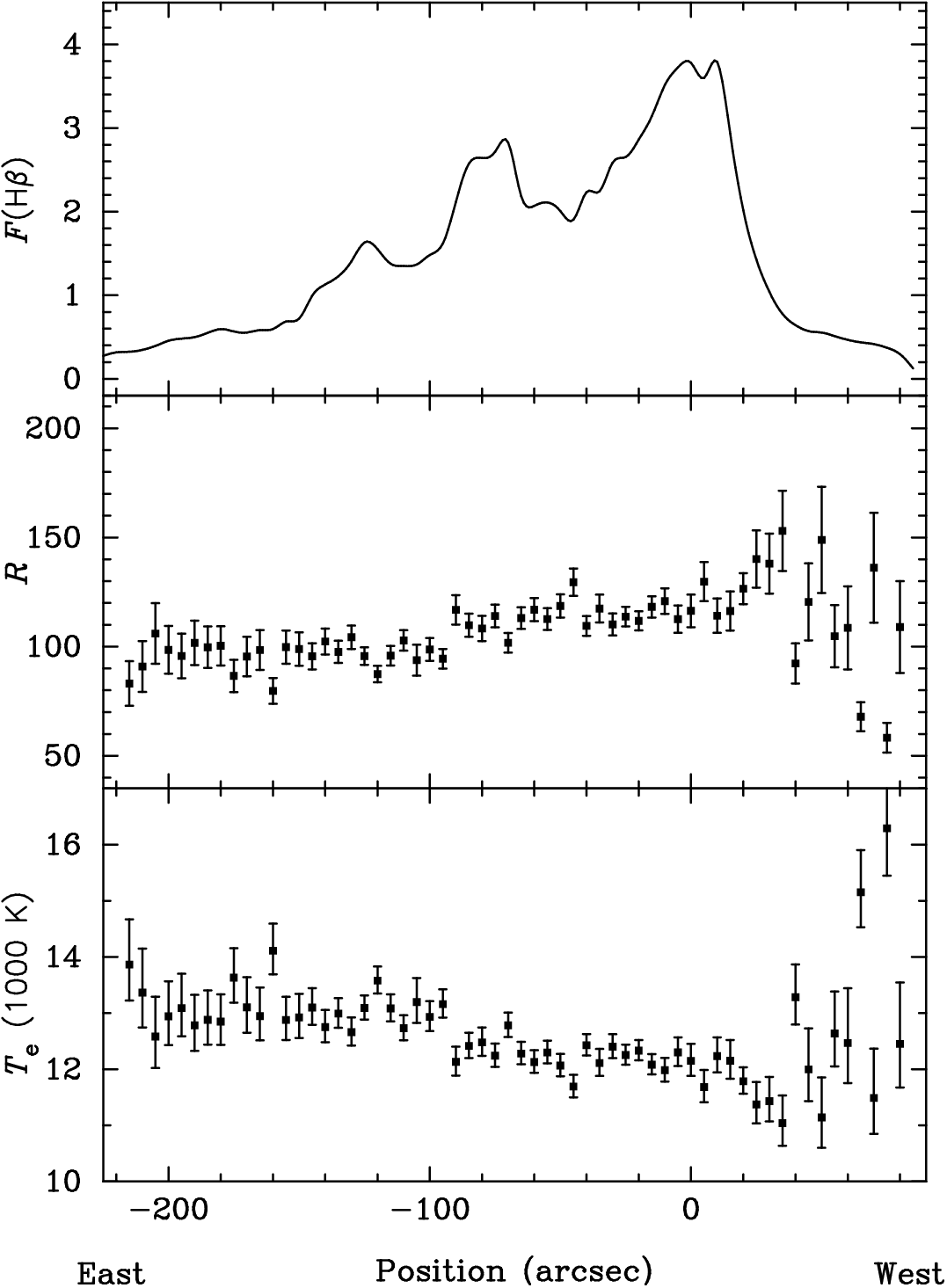}}
\caption{Spacial profile of the H$\beta$ flux (in the units 10$^{-13}$\,ergs\,cm$^{-2}$\,s$^{-1}$), the [\ion{O}{iii}]($\lambda4659 + \lambda5007)/\lambda4363$ ratio ($R$), and the derived electron temperature along the  $\Delta\delta$ = 30\arcsec N. The position in horizontal axis is relative to the reference star.}
\label{fig:perfilM30}
\end{figure}

The observations were performed on the nights of 8 and 9 November 2002 at the Laborat\'orio Nacional de Astrof\'isica (LNA), Brazil, with the Cassegrain spectrograph attached to the 1.6 m telescope. The detector used was a SITe CCD of 2048 $\times$ 2048 pixels. The spatial scale was 0\farcs56\,pixel$^{-1}$. Using a grid of 1200 grooves mm$^{-1}$, we obtained spectra covering the wavelength range of 4050 \AA{} to 5030 \AA{} with a dispersion of 0.5~\AA{}~pixel$^{-1}$ and resolution of 2.7 \AA. The slit used had an entrance of $360\arcsec\times1\farcs5$ on the plane of the sky and was aligned along the east-west direction. The slit was centered at 3 different declinations relative to the reference star MPG 470 \citep{Massey et al. 1989} ($\alpha=0^{\rm h}59^{\rm m}06^{\rm s}$ and $\delta=-72^\circ 10\arcmin 34\arcsec$, J2000): $\Delta\delta=0\arcsec$, $\Delta\delta=30\arcsec$~N, and $\Delta\delta=30\arcsec$~S. At declination offset $\Delta\delta$ = 30\arcsec~N, the center of slit was moved almost 70\arcsec\ to the east of the reference star to cover a brighter section of \object{NGC\,346}. To limit the effects of cosmic rays, three exposures of 20 minutes were completed at each slit position. At $\Delta\delta=30\arcsec$~S, two additional exposures of 10 minutes were acquired. Several bias and dome flat-fields exposures were taken at the beginning and the end of each night. For wavelength calibration, spectra of He-Ar lamp were taken before and after each object exposure. For flux calibration, the spectrophotometric standard stars \object{HR\,9087} and \object{HR\,1544} were observed.

\setcounter{footnote}{2}

Data reduction was performed using {\it IRAF\footnote{{\it IRAF} is distributed by the National Optical Astronomy Observatory (NOAO), which is operated by the Association of Universities for Research in Astronomy (AURA), Inc., under cooperative agreement with the National Science Foundation.}} software, and the standard procedures for bias correction, flat-fielding, cosmic-ray event removal, and wavelength and flux calibration were followed. From the mean of the two-dimensional spectra obtained at each declination offset $\Delta\delta$, we extracted a series of one-dimensional spectra from contiguous regions in length of 5\arcsec\ along the slit axis. To ensure that independent 1D spectra were extracted from 2D spectra along the given north-south strips, we defined fiducial positions along the slit axis; this was achieved by identifying, inside the 2D spectra spatial profiles, the detected star position and comparing this with the coordinates of this star in direct images of the region. This procedure generated 179 one-dimensional spectra from regions of 5\arcsec$\times$1\farcs5 of size.

The line fluxes were measured by Gaussian fitting of the line profile over a linear local continuum. The measurements were completed with the {\it splot} routine of the {\it IRAF} package. The error associated with the line fluxes were estimated to be $\sigma^2 = \sigma^2_{\rm line} + \sigma^2_{\rm cont}$, where $\sigma_{\rm line}$ is the Poisson error of the emission line and $\sigma_{\rm cont}$ is the error due to the continuum noise, calculated to be $\sigma_{\rm cont} = \sqrt{N} \Delta\sigma_{\rm r.m.s.}$, where $N$ is the number of pixels covered by the emission line, $\Delta$ is the dispersion of the spectrum (units of wavelength per pixel), and $\sigma_{\rm r.m.s.}$ is the root mean square of the continuum flux density (flux per unit wavelength). All line intensities were normalized to H$\beta$ and corrected for the effect of interstellar extinction, by comparing the observed ratios H$\gamma$/H$\beta$ with the theoretical one, calculated by \citet{Storey & Hummer 1995} for an electron temperature of 10\,000 K and a density of 100 cm$^{-3}$. The reddening law for the SMC obtained by \citet{Prevot et al. 1984} was used.

\section{Determination of the electron temperature}
%

Electron temperatures were calculated from the [\ion{O}{iii}]($\lambda4959 + \lambda5007)/\lambda4363$ line intensity ratio by solving numerically the equations of equilibrium for the $n$-level atom ($n=6$) using the {\it temden} routine of the {\it nebular} package of {\it STSDAS/IRAF} \citep[see][]{Shaw & Dufour 1995}. The assumed values for energy levels, transition probabilities and collision strengths were taken from \citet{Bowen 1960}, \citet{Wiese et al. 1996}, and \citet{Lennon & Burke 1994}, respectively. The dependence of the calculated electron temperature $T_{\rm e}$ on the assumed electron density $N_{\rm e}$ is very weak. In our estimates, we adopted a fixed electron density of 100 cm$^{-3}$ as a representative value. Electron densities in the range of 50 to 500 cm$^{-3}$ were found for \object{NGC\,346} \citep{Dufour & Harlow 1977, Peimbert et al. 2000, Tsamis et al. 2003}. We calculated that, for electron densities in this range, the errors in the electron temperature estimates due to uncertainties in density would be below 0.1\%.

\section{Results and Discussions}
%

\begin{figure}
\resizebox{\hsize}{!}{\includegraphics*{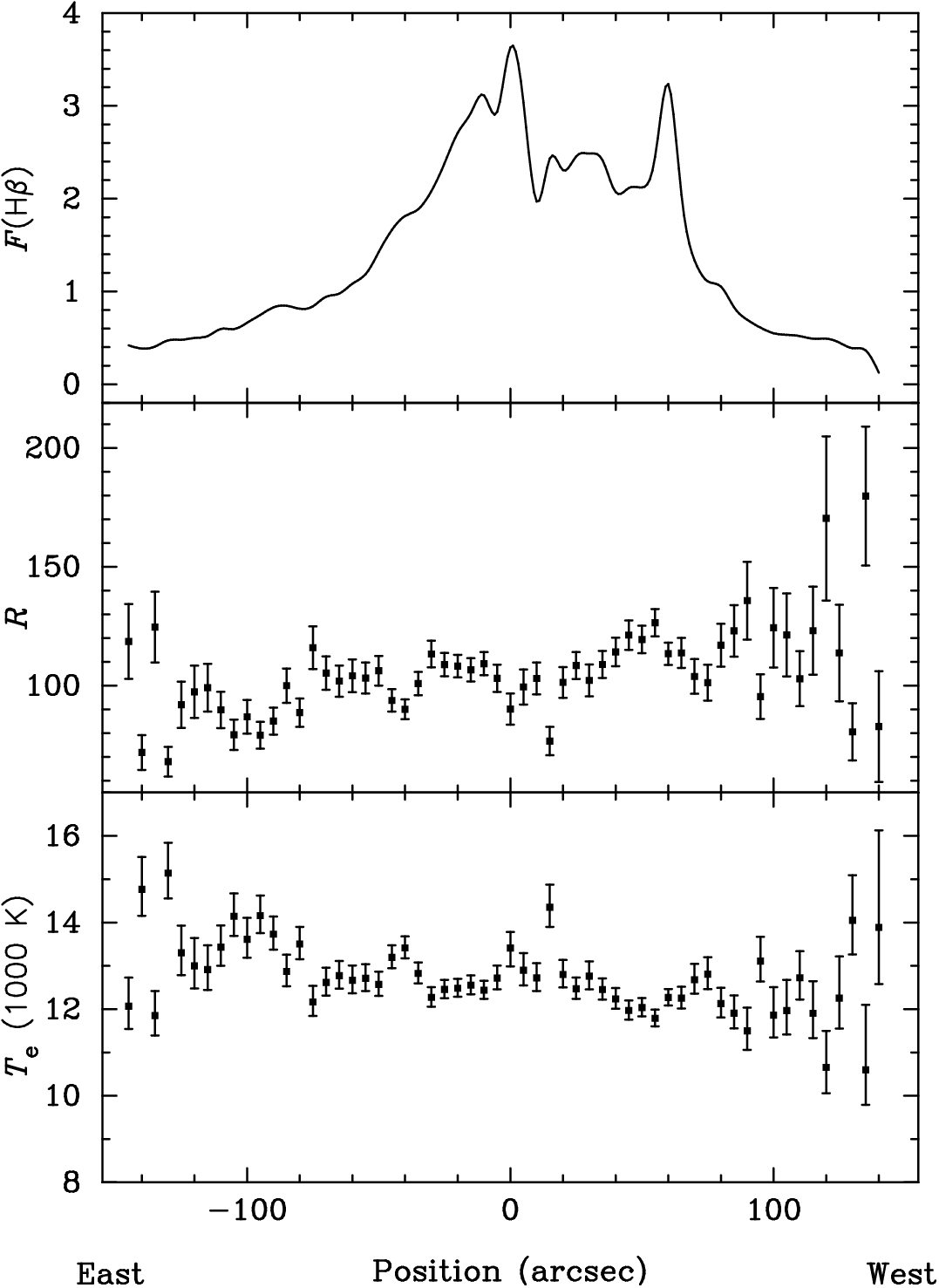}}
\caption{Same as Fig. \ref{fig:perfilM30}, but for $\Delta\delta$ = 0\arcsec.}
\label{fig:perfilZero}
\end{figure}

\begin{figure}
\centering
\resizebox{\hsize}{!}{\includegraphics*{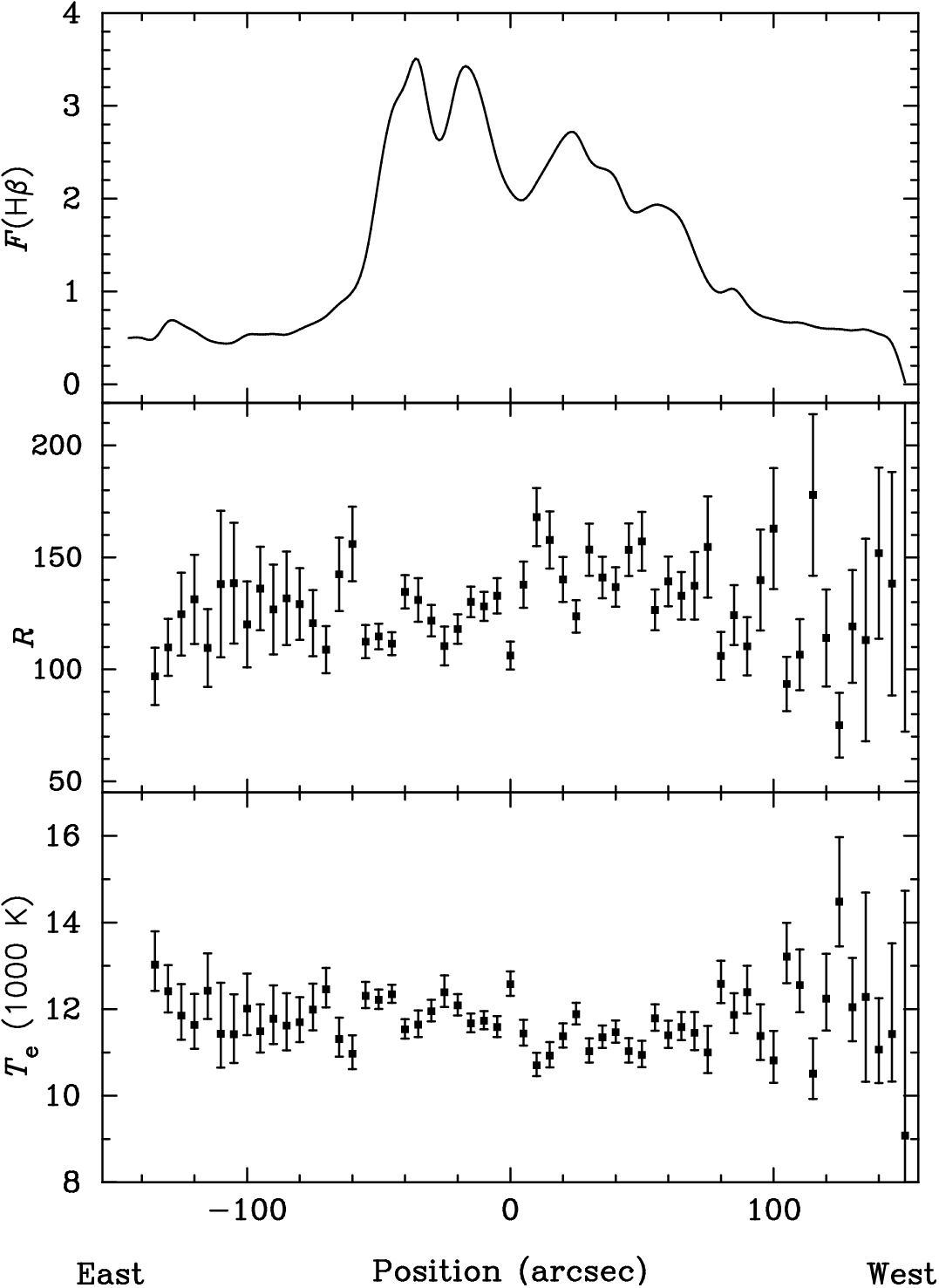}}
\caption{Same as Fig. \ref{fig:perfilM30}, but for $\Delta\delta$ = 30\arcsec S.}
\label{fig:perfilm30}
\end{figure}

Figures \ref{fig:perfilM30}--\ref{fig:perfilm30} show the spatial profiles along the slit of the measurement of H$\beta$ flux, [\ion{O}{iii}] ratio and electron temperature. Table~\ref{tab:estat} presents a summary statistics of the [\ion{O}{iii}] ratio and the inferred electron temperature, including the number $N$ of different nebular sections (one-dimensional spectra), the minimum and maximum, the median, the first quartile $Q1$ (upper limit of the 25\% lowest values) and the third quartile $Q3$ (lower limit of the 25\% highest values). We also indicate both the mean and the standard deviation $\sigma$ weighted by the H$\beta$ flux. The electron temperature estimates are quite homogeneous. The scatter is small with 50\% of the values less than 4.4\% from the  overall median temperature of 12\,331 K. For a perfect Gaussian distribution, the standard deviation is related to the quartiles by $\sigma \approx 1.35 (Q3-Q1)$. Assuming that this relationship holds for the observed temperature distribution, we estimate a dispersion of 6.5\% from the non-parametric statistics. Part of the scatter is real and part is due to measurement errors, as can be inferred by the increase in scatter at the fainter areas at the edges of the nebula. From the parametric statistics (mean and standard deviation weighted by the H$\beta$ flux), which provided more weight to the more precise data, we estimated a weighted mean electron temperature of 12\,269 K and a weighted standard deviation of 6.1\%. The similarity between these two independent estimates of the dispersion in the data indicates that the measurement errors are not dominant sources of scatter in the temperature data. 

\begin{table}[ht]
\caption{Electron temperatures calculated using data from the literature}
\label{tab:te}
\begin{tabular}{rrlrrl}
\hline\hline
\noalign{\smallskip}
Ratio & $T_{\rm e}$ (K) & Ref./Region &  Ratio & $T_{\rm e}$ (K) & Ref./Region \\
\noalign{\smallskip}
\hline
\noalign{\smallskip}
\multicolumn{6}{c}{
[\ion{O}{iii}] $(\lambda4959+\lambda5007)/\lambda4363$ 
} \\
\noalign{\smallskip}
\hline
\noalign{\smallskip}
 99.8 & 12\,875 & [1]      & 105.1 & 12\,625 & [2]      \\
107.4 & 12\,519 & [3]      & 110.9 & 12\,370 & [4] NW   \\
117.9 & 12\,091 & [4] SE   &  96.6 & 13\,044 & [5] I    \\
 96.7 & 13\,040 & [5] II   &  90.1 & 13\,520 & [6] 1-1c \\
 92.6 & 13\,263 & [6] 1-2b & 130.3 & 11\,665 & [6] 2-1c \\
113.3 & 12\,245 & [6] 2-2b &  85.7 & 13\,689 & [7] 1    \\
 92.4 & 13\,279 & [7] 2    & 108.7 & 12\,464 & [7] 3    \\
 98.5 & 12\,942 & [7] 4    &  87.2 & 13\,590 & [7] 5    \\ 
114.6 & 12\,220 & [7] 11   & 113.2 & 12\,275 & [7] 12   \\
111.1 & 12\,363 & [7] 13   & 108.0 & 12\,492 & [7] 14   \\
105.5 & 12\,606 & [7] 15   &  99.1 & 12\,916 & [7] 16   \\
 95.3 & 13\,113 & [7] 17   &  89.1 & 13\,472 & [7] 18   \\
108.2 & 12\,487 & [8] \\
\noalign{\smallskip}
\hline
\noalign{\smallskip}
\multicolumn{6}{c}{
[\ion{N}{ii}] $(\lambda6548+\lambda6583)/\lambda5755$ 
} \\
\noalign{\smallskip}
\hline
\noalign{\smallskip}
 52.22 & 13\,510 & [8] \\
\noalign{\smallskip}
\hline
\noalign{\smallskip}
\multicolumn{6}{c}{
[\ion{O}{ii}] $(\lambda3726+\lambda3729)/(\lambda7320+\lambda7330)$ 
} \\
\noalign{\smallskip}
\hline
\noalign{\smallskip}
40.8 & 11\,627 & [5] I  & 35.5 & 12\,747 & [5] II \\
43.0 & 11\,707 & [7] 1  & 45.6 & 11\,305 & [7] 2  \\
35.4 & 13\,324 & [7] 3  & 35.7 & 13\,262 & [7] 4  \\
45.8 & 11\,263 & [7] 5  & 51.8 & 10\,501 & [7] 11 \\
41.5 & 11\,974 & [7] 12 & 37.7 & 12\,768 & [7] 13 \\
34.0 & 13\,731 & [7] 14 & 24.9 & 17\,902 & [7] 15 \\
32.7 & 14\,125 & [7] 16 & 42.6 & 11\,773 & [7] 17 \\
33.7 & 13\,813 & [7] 18 & 36.7 & 12\,459 & [8]    \\
\noalign{\smallskip}
\hline
\noalign{\smallskip}
\multicolumn{6}{c}{
[\ion{S}{ii}] $(\lambda6716+\lambda6731)/(\lambda4068+\lambda4076)$ 
} \\
\noalign{\smallskip}
\hline
\noalign{\smallskip}
15.2 & 8\,613 & [7] 12 & 15.5 &  8\,525 & [7] 13 \\
14.0 & 9\,110 & [7] 17 & 10.8 & 10\,792 & [8]    \\
\noalign{\smallskip}
\hline
\noalign{\smallskip}
\end{tabular}
\begin{minipage}[t]{\columnwidth}
References:
[1] \citet{Aller & Faulkner 1962}; [2] \citet{Dickel et al. 1964}; [3] \citet{Dufour 1975}; [4] \citet{Dufour & Harlow 1977}; [5] \citet{Peimbert & Torres-Peimbert 1976}; [6] \citet{Pagel et al. 1978}; [7] \citet{Peimbert et al. 2000}; [8] \citet{Tsamis et al. 2003}.
\end{minipage}
\end{table}
 
\begin{figure}[!ht]
\resizebox{\hsize}{!}{\includegraphics*[angle=-90]{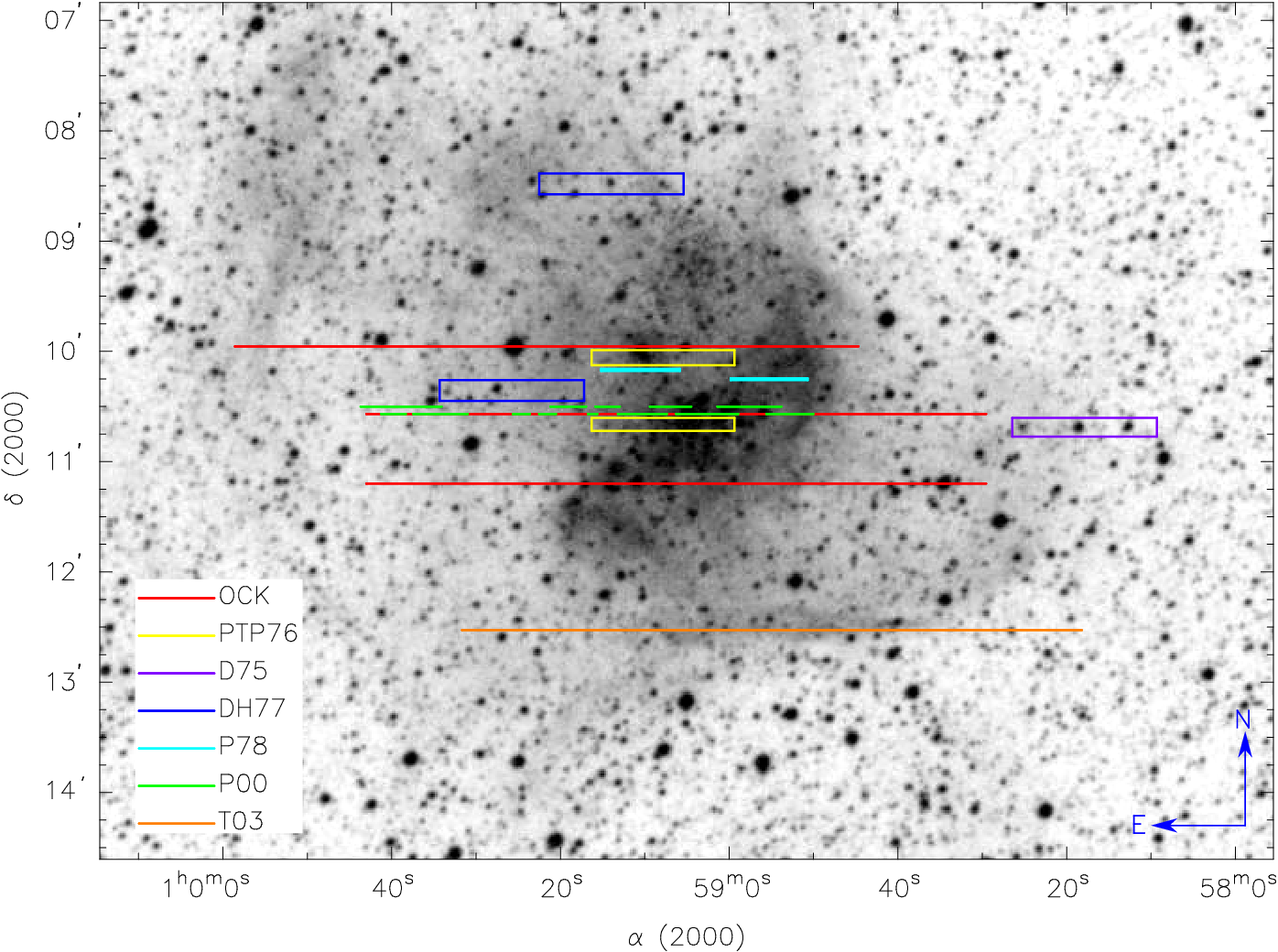}}
\caption{Apertures of the observations of \object{NGC\,346} by: 
(OCK) this paper, 
(D75)   \citet{Dufour 1975}, 
(DH77)  \citet{Dufour & Harlow 1977}, 
(PTP76) \citet{Peimbert & Torres-Peimbert 1976}, 
(P78)   \citet{Pagel et al. 1978}, 
(P00)   \citet{Peimbert et al. 2000}, and 
(T03)   \citet{Tsamis et al. 2003}.}
\label{fig:fendas}
\end{figure}

For comparison with our results, we recalculated the electron temperature from line intensity ratios collected from the literature. Table~\ref{tab:te} presents the temperature diagnostics used, the line ratio values, the electron temperatures recalculated with update atomic parameters, the references for the observational data, and the authors's label for the extraction window for these data, which are shown in Fig.~\ref{fig:fendas}. The temperature estimates obtained from the data in the literature, although corresponding to different areas on the plane of the sky and different ionization zones, are very similar to both each other and our values. For example, the [\ion{O}{iii}] temperatures for data from the literature have a mean of 12\,685 K (3\% higher than our mean value) and a dispersion of only 3.9\%, while the mean [\ion{O}{ii}] temperature is 12\,153 K (1\% lower than our mean value). A significant source of the discrepancies between different temperature estimates is error in the individual measurements, which are at their lowest of the order of 2-3\%. 

\begin{figure}
\resizebox{\hsize}{!}{\includegraphics*[angle=-90]{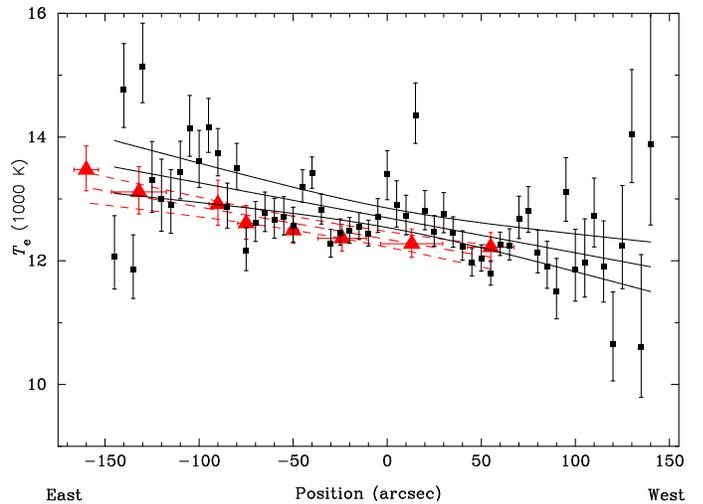}}
\caption{Linear regression analysis for the data along declination offset $\Delta\delta$~=~0\arcsec (squares, solid line) and for the data by \citet{Peimbert et al. 2000} (triangles, dashed line). The curved lines define the 95\% confidence bands for the regression lines.}
\label{fig:linreg}
\end{figure}

All observations support the conclusion that \object{NGC\,346} has a homogeneous temperature structure with variations of about $\pm$6\% or less. It is true that observations, which collect light that has been integrated along the line of sight, tend to smooth out small spatial-scale fluctuations of any line ratio, but localized measurements are able to resolve global internal gradients. Any significant large-scale systematic variation in electron temperature should then be revealed by high signal-to-noise ratio observations, as in studies of the \object{Orion Nebula} \citep{Walter et al. 1992} and the planetary nebulae \object{NGC\,6720} \citep{Garnett & Dinerstein 2001}, \object{NGC\,4361} \citep{Liu 1998}, and \object{NGC\,2438}, \object{NGC\,2440}, \object{NGC\,3132}, \object{NGC\,3242}, \object{NGC\,6302}, and \object{NGC\,7009} \citep{Krabbe & Copetti 2005}. In \object{NGC\,346}, we found, along the declination offsets $\Delta\delta$ = 0\arcsec\ and $\Delta\delta$ = 30\arcsec N, a small but clear electron temperature gradients of $-5.7\pm1.3$ K arcsec$^{-1}$, which is statistically significant at the 95\% confidence level. Using the data obtained by \citet{Peimbert et al. 2000} for areas along a declination offset close to $\Delta\delta$ = 0\arcsec\ (their regions 11-18; see Fig. \ref{fig:fendas}), we found a similar temperature gradient of $-5.2\pm0.8$ K arcsec$^{-1}$. Figures \ref{fig:linreg} shows the result of a linear regression analysis of our data and a comparison with the data of \citet{Peimbert et al. 2000}. Along the declination offset $\Delta\delta$ = 30\arcsec\,S, no significant gradient was found.

\subsection{Magnitude of the electron temperature fluctuations}

The discrepancy between the temperature estimates obtained by means of forbidden line ratios and those derived from the Balmer and radio continuum was explained by \citet{Peimbert 1967} as a consequence of spatial electron temperature fluctuations inside the nebula. To quantify the effects of these fluctuations on the estimated temperature, \citet{Peimbert 1967} introduced the temperature fluctuation parameter $t^2$ defined to be

\begin{equation}
\label{t2}
 t^2 = \frac{\int (T_{\rm e} - T_{\rm 0})^2N_{\rm i}N_{\rm e}\,dV}{T^2_{\rm 0}\int N_{\rm i}N_{\rm e}\,dV} \,,
 \label{eq:t2peimbert}
\end{equation}
with
\begin{equation}
 T_{\rm 0} = \frac{\int T_{\rm e}N_{\rm i}N_{\rm e}\,dV}{\int N_{\rm i}N_{\rm e}\,dV} \,,
 \label{eq:T0}
\end{equation}
where $N_{\rm i}$ is the density of the ion used to measure the temperature, integrations are calculated over the volume V of the nebula. Therefore, the parameters $T_{\rm 0}$, $t^2$, and $\sqrt{t^2}$ are the mean, the relative variance, and the relative standard deviation of the temperature distribution weighted by the square of the local density, respectively. The $t^2$ parameter has also been used to correct the chemical abundances obtained from collisionally excited lines for the effects of temperature fluctuations. Usually, {\it ad hoc} values for $t^2$ are used (including $t^2=0$), or $t^2$ is estimated indirectly from the comparison between temperature values obtained from different indicators.

A direct estimation of $t^2$ can be obtained from localized measurements of the electron temperature across the nebula. We may rewrite Eq.~\ref{t2} and find a lower bound for $t^2$ as
\begin{equation}
\label{t2mod}
t^2=\frac{ \int{\langle(T_\mathrm{e}-T_0)^2\rangle_{\bf\Omega} E_{\bf\Omega} d\Omega}}
             { T_0^2 \int{ E_{\bf\Omega} d\Omega }}\geq
\frac{ \int{(\langle T_\mathrm{e}\rangle_{\bf\Omega}-T_0)^2 E_{\bf\Omega} d\Omega}}
             { T_0^2 \int{ E_{\bf\Omega} d\Omega }} \equiv t^2_\mathrm{sky},
\end{equation}
where: $\langle . \rangle_{\bf\Omega}$ represents the mean value weighted by the square density along the direction
$\bf \Omega$; $E_{\bf\Omega} = \int{N_i N_\mathrm{e} dl}$ is the emission measure; $d\Omega$ and $dl$ are respectively the elements of solid angle and distance along the line of sight. Therefore, the $t^2_\mathrm{sky}$ parameter is the relative variance of the continuous distribution of $\langle T_\mathrm{e}\rangle_{\bf\Omega}$, the mean temperature along the line of sight at the direction $\bf \Omega$ weighted by the emission measure $E_{\bf\Omega}$.

A discrete approximation for $t^2_\mathrm{sky}$, first proposed by \citet{Liu 1998}, is
\begin{equation} \label{ts2}
t_\mathrm{s}^2 \mathrm{(obs)}=\frac{ \sum_{i}{(T_\mathrm{e}^i-T_0)^2 F_i(\mathrm{H}\beta)}}
             { T_0^2 \sum_{i}{F_i(\mathrm{H}\beta)}},
\end{equation}
where $T_\mathrm{e}^i$ and $F_i(\mathrm{H}\beta)$ are the electron temperature and the H$\beta$ flux obtained for the aperture $i$, respectively. This quantity represents the variance (relative to the square of the mean temperature) of the temperature values measured at different apertures, weighted by the H$\beta$ flux. Because part of this variance is due exclusively to errors in the measurements, the final estimation of $t_\mathrm{s}^2$ should be corrected by 
$t_\mathrm{s}^2 = t_\mathrm{s}^2\mathrm{(obs)} - t^{2}_\mathrm{errors},$ where $t^2_{\rm errors}$ is the quadratic mean of the relative errors of temperature, also weighted by the H$\beta$ flux. 

Since the temperature measured for any aperture is the mean value inside a volume crossing the nebula, any temperature variations within distance scales smaller than the aperture size should be smoothed out. Hence, 
$t^2_\mathrm{s}$ is a biased estimator of $t^2$, such that $t^2_\mathrm{s} \leq t^2_\mathrm{sky} \leq t^2$.
However, localized fluctuations in temperatures are unexpected \citep[see][]{Ferland 2001}. Photoionization models of chemically homogeneous nebulae instead predict smooth temperature gradients.

\citet{Copetti 2006} demonstrated with simulation that localized temperature estimates obtained for ions occurring in significant parts of the nebula, such as the [\ion{O}{iii}] and the Balmer jump temperatures, are good tracers of the internal gradient of temperature, and that the values of $t^2_{\rm s}$ calculated from these temperatures contribute significantly to the total variance, with $t^2_{\rm s}(\ion{O}{iii})/t^2 \approx 25\%$ for a typical \ion{H}{ii} region.

For \object{NGC\,346}, we obtained an observed temperature variance of $t^2_{\rm s}({\rm obs}) = 0.00375$ (corresponding to a dispersion of 6.1\%), a mean quadratic error of 4.1\%, and a corrected variance in the temperature distribution on the plane of the sky of  $t^2_{\rm s} = 0.0021$ (or equivalently, a dispersion of 4.5\%). This value is similar to $t^2_{\rm s} = 0.0025$ obtained for the 30 Doradus Nebula \citep{Krabbe & Copetti 2002} and is inside the range of 0.0011 to 0.0050 found in planetary nebulae \citep{Krabbe & Copetti 2005}.

Using the [\ion{O}{iii}] $(\lambda4959+\lambda5007)/\lambda4363$ ratios measured by \citet{Peimbert et al. 2000} for 13 areas of \object{NGC\,346} and recalculating the electron temperature with the same atomic data used in the present paper (see Table~\ref{tab:te}), we calculated a variance in the temperature distribution on the plane of the sky of only $t^2_{\rm s} = 0.00062$, which is even lower than the value that we estimated from our data, possibly because their data extraction windows were up to a factor of 7 larger than ours and were located within a smaller area of \object{NGC\,346}. 

Using the approximation $t^2_{\rm s}(\ion{O}{iii})/t^2 \approx 25\%$ from \citet{Copetti 2006}, we estimate a global temperature fluctuation parameter of $t^2 \approx 0.008$ for \object{NGC\,346}, which is entirely compatible with the values predicted by photoionization models for an object with the characteristics of \object{NGC\,346}, but is too small to explain the abundance discrepancy problem. \citet{Tsamis et al. 2003} determined the O$^{++}$/H$^+$ abundance ratio for \object{NGC\,346} from collisionally excited and recombination lines and found an abundance discrepancy factor of $ADF = 2.3$; this would require significant temperature fluctuations, corresponding to $t^2 = 0.09$, to be described by the temperature fluctuation scenario. Although we cannot exclude the possibility of small spatial-scale temperature fluctuations, since they may not be detected by our localized observations, it is difficult to accept that variations in temperature of dispersion around the mean of the order of 30\% would remain undetected.
 
We emphasize that although the parameter $t^2_{\rm s}(\ion{O}{iii})$ probes the temperature variation in the O$^{++}$ zone, the correction applied to calculate the 3D temperature fluctuation parameter $t^2$ takes into account both the projection effect of the temperature measurements on the plane of the sky and the variation in temperature outside the O$^{++}$ zone. This correction is based on model simulations, and is therefore model dependent. In particular, for a spherical symmetric and constant density nebula, the models by \citet{Copetti 2006} predict values of $t^2_{\rm s}(\ion{O}{iii})/t^2$ of between 8 and 40\% for a wide range of the input parameters (hydrogen density $N_\mathrm{H}$, effective temperature of the ionizing star $T_{\rm eff}$, ionization parameter U). For a more restricted range of parameters ($40\,000 \la T_{\rm eff} \la 50\,000$ K, $N_\mathrm{H} \approx 100-1000$ cm$^{-3}$, $\log U \approx -3$), corresponding to typical \ion{H}{ii} regions like \object{NGC\,346}, a more limited range of ratios of $15\% \le t^2_{\rm s}(\ion{O}{iii})/t^2 \le 35\%$ is found. We therefore expect an error of about 40\% for the estimate of $t^2$ obtained from $t^2_{\rm s}(\ion{O}{iii})$, due solely to the model dependency on the input parameters. Therefore, this estimation of the temperature fluctuation parameter $t^2$, and any other obtained by any method, should always be considered as approximate estimation because of the large associated errors. Even though, it is clear that the temperature fluctuation parameter of $t^2=0.09$ obtained from the comparison between abundances derived from recombination and collisionally excited emission-lines is incompatible with the order of magnitude smaller value of $t^2 = 0.008$ obtained from localized determinations of electron temperature across the nebula.

\citet{Peimbert et al. 2000} inferred a value of $t^2 = 0.022\pm0.012$ for \object{NGC\,346} (and a minimum of $t^2 = 0.0013$) by comparing the mean temperatures measured using different indicators, which estimates the temperature fluctuation that accounts directly for the temperatures differences between different ionization zones. This value is larger but compatible within the error bars with our own estimation. It is still too small however to explain the abundance discrepancy problem.

\section{Conclusions}
%

We have used long-slit spectrophotometric data of high signal-to-noise ratio in the range of 4050 $-$ 5030 \AA\ to study the spatial variation in electron temperature in the \ion{H}{ii} region \object{NGC\,346}. Electron temperatures were derived from the [\ion{O}{iii}]($\lambda4959 + \lambda5007)/\lambda4363$ line ratio for 179 areas of $5\arcsec\times 1\farcs5$ distributed in three different declination. Our main results are the following:
\begin{enumerate}
\item 
A largely homogeneous electron temperature distribution was found with a mean value of 12\,269 K and a standard deviation of 6.1\%, both weighted by the H$\beta$ flux.
   
\item
A small, but statistically significant, large scale gradient in temperature of the order of $-5.7\pm1.3$ K arcsec$^{-1}$ was found along the east-west direction at two different offset positions.

\item
About half of the variance of the temperature was attributed to measurement errors. The remainder corresponded to a temperature fluctuation on the plane of the sky of $t^2_{\rm s} = 0.0021$ (equivalent to a dispersion of 4.5\%), which implied a 3D temperature fluctuation parameter of $t^2 \approx 0.008$. 

\item
The magnitude of the temperature fluctuations observed is in agreement with the large scale variations in temperature predicted by standard photoionization models, but is too low to explain the abundance discrepancy problem. However, the existence of small spatial scale temperature variations is not excluded by the present observations.
\end{enumerate} 

\begin{acknowledgements}
We thank the anonymous referee for helpful comments and suggestions.
We wish to thank the staff of the Laborat\'{o}rio Nacional de Astrof\'{\i}sica for their assistance during the observations. This work was supported by the Brazilian agencies CAPES and CNPq.
\end{acknowledgements}

\end{document}